\documentclass[apj]{emulateapj}

\newcommand{\DDir}{\relax{D\kern-.7em{/}}}

\newcommand{\be}{\begin{equation}}
\newcommand{\ee}{\end{equation}}
\newcommand{\bea}{\begin{equation*}}
\newcommand{\eea}{\end{equation*}}

\newcommand{\nin}{\relax{\in\kern-.8em{/}}}

\newcommand{\keV}{\mbox{ keV}}

\newcommand{\TeV}{\mbox{ TeV}}

\usepackage{amsmath}

\begin{document}

\title{Hard X-ray emission from accretion shocks around galaxy clusters}

\author{Doron Kushnir\altaffilmark{1} and
Eli Waxman\altaffilmark{1}}
\altaffiltext{1}{Physics Faculty,
Weizmann Institute of Science, Rehovot, Israel}

\begin{abstract}
We show that the hard X-ray (HXR) emission observed from several galaxy clusters is consistent with a simple model, in which the nonthermal emission is produced by inverse Compton scattering of cosmic microwave background photons by electrons accelerated in cluster accretion shocks: The dependence of HXR surface brightness on cluster temperature is consistent with that predicted by the model, and the observed HXR luminosity is consistent with the fraction of shock thermal energy deposited in relativistic electrons being $\lesssim0.1$. Alternative models, where the HXR emission is predicted to be correlated with the cluster thermal emission, are disfavored by the data. The implications of our predictions to future HXR observations (e.g. by NuStar, Simbol-X) and to (space/ground based) $\gamma$-ray observations (e.g. by Fermi, HESS, MAGIC, VERITAS) are discussed.
\end{abstract}




\section{Introduction}
\label{sec:Introduction}

Nonthermal emission is observed from several clusters of galaxies, mainly in the radio band \citep[e.g.,][]{feretti2008cgr}. In some cases, nonthermal hard ($>20\,\textrm{keV}$) X-ray (HXR) emission is also observed \citep[for review, see][]{rephaeli2008npc}. The radio emission is interpreted as synchrotron radiation, thereby suggesting that relativistic electrons and magnetic fields are present in the intracluster medium (ICM). The HXR emission is usually interpreted as due to inverse Compton (IC) scattering of cosmic microwave background (CMB) photons by nonthermal relativistic electrons \citep[e.g.,][]{rephaeli1979rei,sarazin1999esp}. In some cases, however, the HXR emission is also consistent with a two-temperature ICM plasma \citep[see][and refrences therein]{rephaeli2008npc}.

An accurate determination of the nonthermal HXR flux requires a precise measurement of the thermal emission, which dominates up to $\sim30-40\,\textrm{keV}$. This, as well as a clear answer to the question of whether the HXR emission is due to a two-temperature plasma or to the presence of a non-thermal electron population, will only be provided by future HXR and $\gamma$-ray missions. In this paper, we examine the consequences of the nonthermal HXR detections reported in the literature, assuming that the reported fluxes are correct (i.e. that the substraction of the thermal component has been preformed correctly) and that the energy distribution of the nonthermal photons is well described by a power-law (and is not due, therefore, to a two-temperature plasma).
We consider all reported measurements of HXR emission, excluding those where confusion with AGNs or radio galaxies (located within the field of view of the cluster observation) is possible (At present, data are available for 13 clusters satisfying this criterion, see \S~\ref{sec:observations} for details).

Several models for the HXR emission from clusters have been presented in the literature. These models differ in the assumptions regarding the nonthermal emission mechanism as well as regarding the origin of the emitting electrons. In some models, the nonthermal emission mechanism is IC scattering of CMB photons by relativistic electrons \citep[e.g.,][]{rephaeli1979rei,sarazin1999esp,colafrancesco2009ics}, while in others the mechanisms are nonthermal bremsstrahlung \citep[e.g.,][]{sarazin1999esp,sarazin2000nbh} or synchrotron emission from ultra-relativistic electrons \citep{timokhin2004ont,inoue2005hxr,eckert2007pas}. Various sources have been suggested for the emitting electrons: a population of point sources \citep[e.g. AGN as in][]{katz1976oxr,fabian1976gcs,fujita2007nea}, merger shocks \citep[e.g.,][]{fujita2003nep,brunetti2004arr}, dark matter bow shocks \citep[e.g.,][]{bykov2000nec}, ram-pressure stripping of infalling galaxies \citep[e.g.,][]{deplaa2006ces} and accretion shocks \citep[e.g.,][]{loeb2000cgr,fujita2003nep,berrington2003npa,gabici2003nrc,brunetti2004arr,inoue2005hxr,kushnir2009non}.

We examine in this paper the consistency of the predictions of different models for the HXR emission with available HXR observations. Two types of models are considered. In models of the first type, the HXR emission is predicted to be correlated with the cluster thermal emission, and therefore to be strongly concentrated towards the cluster's center. Models of this type are largely motivated by the spatial correlation observed in some clusters between the nonthermal radio emission and the thermal X-ray emission. A widely discussed model of this type is a model where the HXR emission is due to secondary electrons produced by inelastic p-p collisions between cluster cosmic-rays (CRs) and thermal intra-cluster gas \citep[e.g.,][]{dennison1980frh}. In this model, the nonthermal radio emission and the HXR emission are produced by the same population of electrons. In the second type of models, the nonthermal HXR emission originates in the cluster accretion shocks, and is therefore extended across the cluster. In such models the radio and HXR radiation are produced by different electron populations.

The cluster sample that we use is described in \S~\ref{sec:observations}. In \S~\ref{sec:simple} we show that all available HXR cluster observations are consistent with a simple analytic model \citep[][]{loeb2000cgr,kushnir2009non}, in which the HXR emission is due to IC scattering of CMB photons by relativistic electrons accelerated in accretion shocks surrounding the clusters
\citep[\S~\ref{sec:simple} includes a brief description of the model. The reader is referred to ][for a detailed description]{kushnir2009non}. In \S~\ref{sec:models} we show that models, in which the HXR emission is predicted to be correlated with the cluster thermal emission, are disfavored by the data. This conclusion is based on two main findings. First, we show that Swift's upper limits on HXR emission from several clusters are difficult to explain in models where the emission is dominated by the cluster core (while naturally explained in the accretion model). Second, we show that in models where HXR emission is due to secondary electrons produced by inelastic p-p collisions, the energy in CR protons is required to exceed the thermal energy of the gas in order to explain the detected fluxes. This is both unlikely and inconsistent with the observed correlation between the radio flux and the thermal X-ray flux \citep{kushnir2009mfc}. Our results are summarized and discussed in \S~\ref{sec:conclusions}, with emphasis on predictions of the accretion model that discriminate it from other models and which may be tested by future observations.

The following point should be clarified here. It is difficult to directly determine using current data whether the observed HXR emission is extended or dominated by the cluster core. This is due mainly to the fact that imaging information in the HXR band is at best limited\footnote{One exception is the INTEGRAL measurements of HXR emission from the Coma cluster, which indicate a source extended well beyond the cluster core \citep{eckert2007swe,lutovinov2008xro}. A firm conclusion can not, however, be drawn, since the statistical significance of the Coma HXR detection by INTEGRAL is low.}. We show, however, that this difficulty may be partially overcome, and that information regarding the spatial distribution of the HXR emission may be obtained, by comparing  the fluxes measured by different instruments which differ in their fields of view (FOV).

The FOV radii of RXTE, BeppoSAX and INTEGRAL are in the range of 30' to 60', and those of Swift/BAT and Suzaku are smaller, $\sim10'$. Since these FOV radii are comparable to the characteristic sizes of massive clusters (lying at distances of few 100~Mpc), the HXR flux measurements obtained by instruments with differing FOVs provide some information on the intra-cluster spatial distribution of the HXR emission. Models in which the HXR emission follows the thermal emission predict that the HXR flux should not significantly vary as the FOV grows to include cluster regions beyond the cluster core, since the thermal emission is strongly dominated by the core. On the other hand, models in which the HXR surface brightness is roughly uniform across the cluster, or  rising away from the cluster center (as predicted by the model described in \S~\ref{sec:simple}), predict that the HXR flux should increase significantly as the FOV grows beyond the angular size of the core. As we show in \S~\ref{sec:models}, the data support the latter qualitative behavior, therefore suggesting that the HXR emission is extended and not dominated by the cores of the clusters. It should, however, be kept in mind that comparing the fluxes measured by different instruments is subject to uncertainties, since different instruments may be subject to different systematic effects \citep[see, e.g.,][]{rossetti2004ith,rossetti2007tcc}. Thus, only future HXR (and $\gamma$-ray) missions, capable of producing high resolution maps of clusters, would provide a clear determination of the spatial distribution of the HXR emission (see \S~\ref{sec:conclusions} for discussion).

Throughout, a $\Lambda$CDM cosmological model is assumed with $H_{0}=70h_{70}\,\textrm{km}\,\textrm{s}^{-1}\,\textrm{Mpc}^{-1}$, $\Omega_{m}=0.23$, $\Omega_{b}=0.039$ and $\Omega_{\Lambda}=1-\Omega_{m}$. Due to the small redshift range of the observed clusters, we neglect redshift dependencies where justified.


\section{Observations}\label{sec:observations}

We use the compilation of \citet{rephaeli2008npc} for the RXTE, BeppoSAX and INTEGRAL observations, the results of \citet{ajello2009gcs} for the Swift/BAT observations and the results of \citet{wik2009ssn} for the Suzaku observations. We consider $6$ clusters observed with RXTE (in the $20-80\,\textrm{keV}$ band), $4$ clusters observed with BeppoSAX ($20-80\,\textrm{keV}$), $1$ cluster observed with INTEGRAL ($44-107\,\textrm{keV}$), $8$ clusters observed with Swift/BAT ($50-100\,\textrm{keV}$) and $1$ cluster observed with Suzaku ($12-70\,\textrm{keV}$), where the Coma cluster has been observed with all five instruments, $A2319$ with three and $A2256$ with two. Our sample includes therefore a total of $13$ different clusters, for which HXR detection or upper limits are available. Most of the clusters show clear signs of a recent merger (this may bias the inferred model parameter values, see discussion in \S~\ref{sec:conclusions}). A list of all the clusters in our sample, including their relevant properties, is given in table~\ref{tbl:cluster params}.

\begin{deluxetable*}{ccccccccc}
\tablecaption{The properties of clusters included in our sample$^1$  \label{tbl:cluster params}}
\tablewidth{0pt} \tablehead{ \colhead{Cluster name} & \colhead{$T\,[\textrm{keV}]$} & \colhead{$\beta$} & \colhead{$z$} & \colhead{$r_{200}\,[h_{70}^{-1}\,\textrm{Mpc}]$} & \colhead{$r_{c}\,[h_{70}^{-1}\,\textrm{kpc}]$} & \colhead{$L_{X}\,[10^{45}\,h_{70}^{-2}\,\textrm{erg}\,\textrm{s}^{-1}]^2$} &
\colhead{HXR Flux $[10^{-12}\,\textrm{erg}\,\textrm{cm}^{-2}\,\textrm{s}^{-1}]$}} \startdata Coma$^a$ & $8.38^{+0.34}_{-0.34}$ & $0.654^{+0.019}_{-0.021}$ & 0.0232 & 2.30 & 246 & 1.14 & $21\pm6^{\rm R}$ \\
 &  &  &  &  &  & &  $15\pm5^{\rm B}$ \\
  &  &  &  &  &  & & $18\pm11^{\rm I}$ \\
   &  &  &  &  &  & & $1.7\pm0.17^{\rm Sw}$ \\
    &  &  &  &  &  & & $<10^{\rm Sz}$ \\
A2319$^a$ & $8.8\pm0.5$ & $0.591^{+0.013}_{-0.012}$ & 0.0564 & 2.26 & 204 & 2.44 & $14\pm3^{\rm R}$ \\
 &  &  &  &  &  & & $<23^{\rm B}$ \\
  &  &  &  &  &  & & $<0.67^{\rm Sw}$ \\
A2256$^a$ & $6.6\pm0.4$ & $0.914^{+0.054}_{-0.047}$ & 0.0601 & 2.40 & 419 & 1.16 & $4.6\pm2.4^{\rm R}$ \\
 &  &  &  &  &  & & $8.9^{+4.0}_{-3.6}$~$^{\rm B}$ \\
A2163$^a$ & $13.29\pm0.64$ & $0.796^{+0.03}_{-0.028}$ & 0.201 & 3.21 & 371 & 6.62 & $11^{+17}_{-9}$~$^{\rm R}$ \\
A3667$^a$ & $7\pm0.6$ & $0.541\pm0.008$ & 0.056 & 1.92 & 199 & 1.24 & $<4^{\rm R}$ \\
1ES0657-55.8$^b$ & $17.4\pm2.5$ & $0.62\pm0.007$ & 0.296 & 3.22 & 257 & 7.14 & $5\pm3^{\rm R}$ \\
A2199$^a$ & $4.1\pm0.08$ & $0.655^{+0.019}_{-0.021}$ & 0.0302 & 1.62 & 99.3 & 0.404 & $9.8\pm4^{\rm B}$ \\
A3266$^a$ & $8\pm0.5$ & $0.796^{+0.02}_{-0.019}$ & 0.0594 & 2.46 & 403 & 1.22 & $<0.57^{\rm Sw}$ \\
A3571$^a$ & $6.9\pm0.2$ & $0.613\pm0.01$ & 0.0397 & 2.04 & 129 & 1.04 & $1.4\pm0.5^{\rm Sw}$ \\
A2029$^a$ & $9.1\pm1$ & $0.582\pm0.004$ & 0.0767 & 2.29 & 59.3 & 2.62 & $<1.27^{\rm Sw}$ \\
A2142$^a$ & $9.7^{+1.5}_{-1.1}$ & $0.591\pm0.006$ & 0.0899 & 2.36 & 110 & 3.36 & $<1.50^{\rm Sw}$ \\
Triangulum$^a$ & $9.6\pm0.6$ & $0.61\pm0.01$ & 0.051 & 2.39 & 199 & 1.96 & $<0.65^{\rm Sw}$ \\
Ophiuchus$^a$ & $10.26\pm0.32$ & $0.747^{+0.035}_{-0.032}$ & 0.028 & 2.73 & 199 & 1.95 & $<2.80^{\rm Sw}$ \\
\enddata
\tablenotetext{1}{Error bars given when relevant to our analysis.}
\tablenotetext{2}{Bolometric luminosity}
\tablenotetext{a}{Data taken from \citet{reiprich2002mfx}}
\tablenotetext{b}{Data taken from \citet{tucker1998acf}}
\tablenotetext{B}{BeppoSAX measurements}
\tablenotetext{I}{INTEGRAL measurements}
\tablenotetext{R}{RXTE measurements}
\tablenotetext{Sw}{Swift/BAT measurements}
\tablenotetext{Sz}{Suzaku measurements}
\end{deluxetable*}

Different instruments are sensitive over different X-ray energy ranges. When comparing the measurements of different instruments we therefore compare the flux per logarithmic energy interval of X-ray photons, i.e. the flux divided by $\Lambda\equiv\log(\nu_{\rm max}/\nu_{\rm min})$ where $h\nu_{\rm max,min}$ are the upper and lower bounds respectively of the instrument's energy band. For the relevant instruments, $\Lambda$ is within the range $0.7-1.8$. Our choice is motivated by the fact that a photon spectral index of 2, i.e. $dn_\gamma/d\varepsilon_\gamma\propto\varepsilon_\gamma^{-2}$, is expected in many models (see \S~\ref{sec:simple} \& \S~\ref{sec:models}) and is also consistent with observations \citep[note, however, that there are large uncertainties in the observational determination of the spectral index, e.g.][]{rephaeli2008npc}. For such a spectrum, the flux per logarithmic energy interval is independent of energy.


\section{A simple model for the HXR emission}\label{sec:simple}

Let us first briefly describe the main assumptions, and the main relevant results, of the model discussed by \citet{kushnir2009non} for the nonthermal emission produced by cluster accretion shocks. In this model, it is assumed that matter is accreted onto a cluster of mass $M$ at a rate $\dot{M} =f_{\textrm{inst}} M_{200}/t_H$, where $t_H$ is the (instantaneous) Hubble time, $M_{200}$ is the mass contained within a radius $r_{200}$, within which the mean density is $200$ times the critical density $\rho_{\textrm{crit}}$, and $f_{\textrm{inst}}$ is a dimensionless parameter of order unity, reflecting the temporal fluctuations of $\dot{M}/(M_{200}/t_H)$. As discussed in \citet{kushnir2009non}, 3D numerical simulations indicate that the average value of $f_{\textrm{inst}}$ is $\approx0.5$. The accreted gas is assumed to be shocked to the cluster's virial temperature $T$ (see discussion in \S~\ref{sec:conclusions}). Since the accretion shock is strong and collisionless, it is assumed that it produces a nonthermal population of relativistic electrons, with an energy spectrum \citep[e.g.,][]{blandford1987paa}
\begin{eqnarray}\label{eq:flat spectrum}
dn/d\varepsilon\propto \varepsilon^{-2}.
\end{eqnarray}
The fraction of the post shock thermal energy density carried by relativistic electrons is denoted by $\eta_{e}$.

The accelerated electrons lose energy by IC scattering of CMB photons (which dominates over synchrotron emission at the accretion shock). At sufficiently high energies, where the electron cooling time is short compared to the time scale for cluster evolution, the resulting IC luminosity per logarithmic photon energy interval is simply given by
\begin{eqnarray}\label{eq:shock IC gen}
\nu L_{\nu}^{\textrm{IC,shock}} &=&
\frac{1}{2}\frac{3}{2}\frac{\eta_{e}f_{b}T}{\Lambda_{e}}f_{\textrm{inst}} \frac{M_{200}}{\mu m_{p}t_H},
\end{eqnarray}
where $f_{b}=\Omega_{b}/\Omega_{m}$, $\mu m_p$ is the average mass of shocked plasma particles, and $\Lambda_{e}\sim20$ is the number of logarithmic energy intervals in the energy spectrum of the relativistic electrons. The frequency at which an electron emits most of its IC power is given by $\nu=\nu_{0}\gamma^{2}$, where $\nu_{0}=3T_{\textrm{CMB}}/h$, $T_{\textrm{CMB}}$ is the CMB temperature and $\gamma$ is the Lorentz factor of the electron. The Lorentz factor of electrons emitting HXR photons of energy $\varepsilon_{\textrm{HXR}}$ is
\begin{eqnarray}
&\gamma_{\textrm{HXR}}\approx 5.3\times10^3\left(\frac{\varepsilon_{\textrm{HXR}}}{20\,\keV}\right)^{1/2},
\end{eqnarray}
and their cooling time is
\begin{eqnarray}\label{eq:t_cool}
&t_{\textrm{cool}}\approx 0.44\left(\frac{\varepsilon_{\textrm{HXR}}}{20\,\keV}\right)^{-1/2}\,\textrm{Gyr}.
\end{eqnarray}
Thus, the cooling time of electrons emitting at the HXR band is short compared to the clusters' evolution time, and the IC luminosity at the HXR band is well approximated by eq.~(\ref{eq:shock IC gen}).

In order to determine the cluster's HXR luminosity (and surface brightness) as function of its temperature $T$, a relation between $T$ and $M_{200}$ (and $r_{200}$) should be used. The density profile of the X-ray emitting ICM is generally well described by a "$\beta$-model" \citep{cavaliere1976xrh,gorenstein1978sxr,jones1984scg},
\begin{equation}\label{eq:beta model}
\rho_{\textrm{gas}}(r)\propto\left(1+\frac{r^{2}}{r_{c}^{2}}\right)^{-(3/2)\beta},
\end{equation}
where $r_c$ is the X-ray core radius. Assuming the ICM plasma to be isothermal and in hydrostatic equilibrium gives
\begin{eqnarray}\label{eq:r200,M200}
r_{200}&\sim&\left(\frac{800\pi}{3}\rho_{\textrm{crit}}\right)^{-1/2} \left(\frac{3\beta T}{\mu m_{p}G}\right)^{1/2} \nonumber \\ &\sim&3.1\beta^{1/2}T_{1}^{1/2}h_{70}^{-1}\,\textrm{Mpc},\nonumber \\
M_{200}&\sim&\left(\frac{800\pi}{3}\rho_{\textrm{crit}}\right)^{-1/2} \left(\frac{3\beta T}{\mu m_{p}G}\right)^{3/2} \nonumber \\ &\sim&3.5\cdot10^{15}\beta^{3/2}T_{1}^{3/2}h_{70}^{-1}M_{\odot}.
\end{eqnarray}
Here, $T_{1}=T/10\,\textrm{keV}$ and $\mu\sim0.59$ is the mean molecular weight for fully ionized gas with hydrogen mass fraction of $\chi=0.75$. We assume that the accretion shock is located at $r\sim r_{200}$, since spherical collapse models predict a cluster virial density $<\rho_{\textrm{vir}}>\simeq178\rho_{\textrm{crit}}$ for $\Omega_{m}=1$, $\Omega_{\Lambda}=0$ (with weak dependance on the background cosmology for the relevant range $0.3\lesssim\Omega_{m}<1$).

In addition to the assumption that the ICM is isothermal and in hydrostatic equilibrium, we have assumed in deriving eq.~\eqref{eq:r200,M200} that the density profile is given by eq.~(\ref{eq:beta model}) out to $r_{200}$. This implies $\rho\propto r^{-2}$ at large radii, in contrast with the $\rho\propto r^{-3}$ dependence expected at large $r$ \citep[e.g.][]{navarro1997udp}. However, a detailed discussion of the accuracy of cluster mass determination under these approximations, given in \citet{reiprich2002mfx}, shows that eq.~\eqref{eq:r200,M200} may overestimate $M_{200}$ by no more than $20\%$.

Since most of the clusters observed in HXR have recently undergone a merger, a note is in place concerning the validity of our approximate description of cluster properties (eqs.~\ref{eq:shock IC gen}, \ref{eq:beta model}, \ref{eq:r200,M200}). X-ray cluster maps \citep[e.g.,][]{finoguenov2005xmm} show that both relaxed and unrelaxed clusters are not simple hydrostatic equilibrium systems. Particularly, the structure of recently merged clusters often deviates from a hydrostatic equilibrium. However, since the deviations of the gas profiles from the mean hydrostatic equilibrium profiles are at the few tens of percent level, we expect our description to be approximately valid. Detailed analysis of numerical simulations of cluster mergers \citep[e.g.][]{ricker2001ofa,poole2006imr,poole2007imr} can be used to determine the effects of deviations from our simple approximate description.

Using eq.~\eqref{eq:r200,M200}, the accretions shock luminosity, eq.~(\ref{eq:shock IC gen}), gives
\begin{eqnarray}\label{eq:shock IC app}
\nu L_{\nu}^{\textrm{IC,shock}} &=& 1.7\cdot10^{44}\left(f_{\textrm{inst}}\eta_{e}\right)_{-1}\beta^{3/2}\nonumber\\
&\times&\left(\frac{f_{b}}{0.17}\right) T_{1}^{5/2} \,\textrm{erg}\,\textrm{s}^{-1},
\end{eqnarray}
where $\left(f_{\textrm{inst}}\eta_{e}\right)_{-1}=f_{\textrm{inst}}\eta_{e}/10^{-1}$. Assuming that the HXR emission originates in a thin shell lying at the shock radius, the HXR flux of a cluster at a distance $d\simeq cz/H_{0}$, within the energy band $[\varepsilon_{1},\varepsilon_{2}]$ and within a disk of angular radius $\theta$ centered at the cluster center, is given by
\begin{eqnarray}\label{eq:S for observations shockIC}
F_{[\varepsilon_{1},\varepsilon_{2}]}(\theta) &=&
7.5\cdot10^{-8}\left(\langle f_{\textrm{inst}}\rangle_{\theta}\eta_{e}\right)_{-1}\beta^{1/2} \nonumber \\ &\times& \left(\frac{f_{b}}{0.17}\right) T_{1}^{3/2} \Lambda g_{\rm acc.}(\theta) h_{70}^2 \,\frac{\textrm{erg}}{\textrm{cm}^{2}\textrm{s}},
\end{eqnarray}
where
\begin{eqnarray}\label{eq:g_acc}
g_{\rm acc.}(\theta)= 2\theta_{200}^{2}\left[1-\sqrt{1-\left(\frac{\theta}{\theta_{200}}\right)^{2}}\right].
\end{eqnarray}
Here $\theta_{200}=r_{200}/d$ and we have used $\langle f_{\textrm{inst}}\rangle_{\theta}$, defined as the average value of $f_{\textrm{inst}}$ over the disk considered, to explicitly reflect the possible spatial dependence of the accretion mass flux.

Eq~(\ref{eq:g_acc}) gives an approximate description of the dependence of $F$ on $\theta$, for the case where the emission takes place within a shell of radius $r=r_{200}$ and thickness $w\ll r_{200}$. The thickness of the emitting region is approximately given by the product of the cooling time of the emitting electrons and the velocity of the downstream fluid relative to the shock velocity, $u_{d}$. Using $u_{d}=\sqrt{T/3\mu m_{p}}\simeq7.4\cdot10^{2}T_{1}^{1/2}\,\textrm{km}\,\textrm{s}^{-1}$, we have
\begin{eqnarray}\label{eq:thickness}
w&\simeq& u_{d}t_{cool} \nonumber\\
&\simeq& 3.3\cdot10^{2} T_{1}^{1/2} \left(\frac{\varepsilon_{\textrm{HXR}}}{20\,\keV}\right)^{-1/2}\,\textrm{kpc}.
\end{eqnarray}
This confirms that $w\ll r_{200}$. For $w=0.1r_{200}$, eq.~(\ref{eq:g_acc}) is accurate to better than $\sim25\%$ for any $\theta$.

\begin{figure}
\epsscale{1.2} \plotone{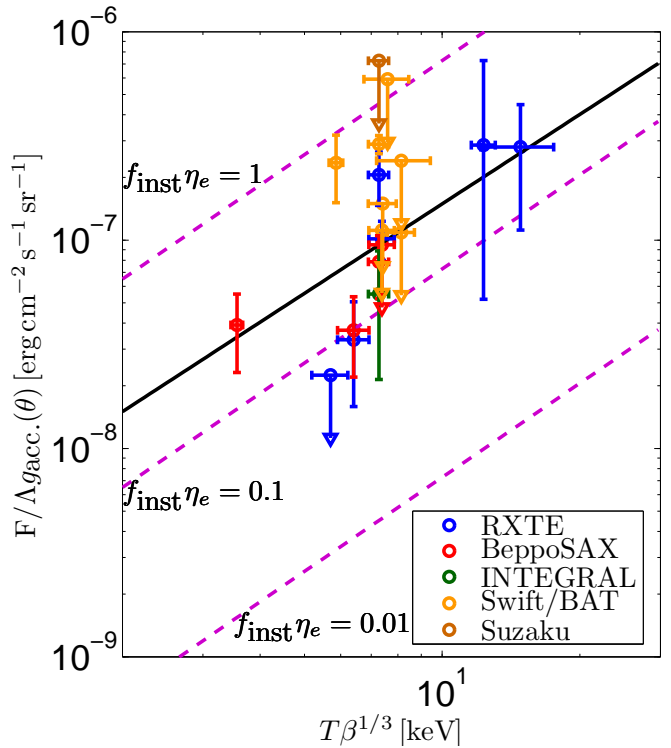} \caption{$F/\Lambda g_{\rm acc.}$ as function of $T\beta^{1/3}$ (see eq.~\ref{eq:S for observations shockIC}). A linear fit for $\ln(F/\Lambda g_{\rm acc.})$ as function of $\ln(T\beta^{1/3})$ gives a slope of $1.4\pm0.5$, consistent with the predicted slope of $3/2$. For a slope of $3/2$, the best linear fit is obtained for $f_{\textrm{inst}}\eta_{e}\sim0.2$. In deriving the fits, we used average values (over all instruments) for the Coma and $A2256$ fluxes, and disregarded the Swift/BAT measurement, for which the existence of a nonthermal component is uncertain \citep{ajello2009gcs}. The best fit is shown by the solid black line. Constant $f_{\textrm{inst}}\eta_{e}$ lines are shown as dashed magenta lines.
\label{HXR}}
\end{figure}

\begin{figure}
\epsscale{1.2} \plotone{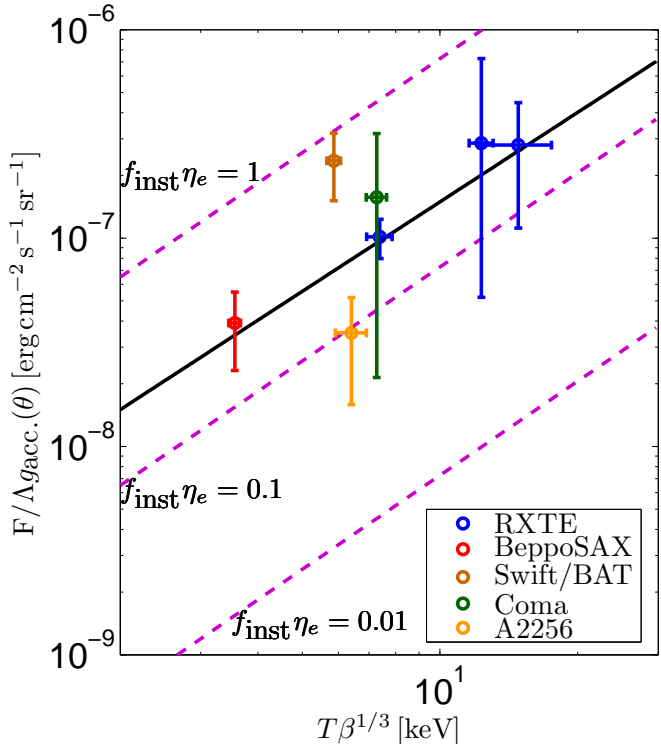} \caption{Same as figure~\ref{HXR}, including only clusters with HXR detections, and average (over all instruments) values for the Coma and $A2256$ clusters.
\label{HXR_noup_cs}}
\end{figure}

Eq.~\eqref{eq:S for observations shockIC} predicts that $F/\Lambda g_{\rm acc.}$ should scale as $(T_{1}\beta^{1/3})^{3/2}$, with a normalization that depends on $\langle f_{\textrm{inst}}\rangle_{\theta}\eta_{e}$. The model predictions are compared with observations in figure~\ref{HXR}, where we show the measurements and upper-limits of $F/\Lambda g_{\rm acc.}$ for the entire sample, and in figure~\ref{HXR_noup_cs}, where we show only clusters with HXR detections (The vertical error bars reflect only the uncertainty in measured fluxes, $F$; we ignored the errors in the determination of $g_{\rm acc.}$, which are significantly smaller). A linear fit for $\ln(F/\Lambda g_{\rm acc.})$ as function of $\ln(T\beta^{1/3})$ gives a slope of $1.4\pm0.5$, consistent with the predicted slope of $3/2$. For a slope of $3/2$, the best linear fit is obtained for $f_{\textrm{inst}}\eta_{e}\sim0.2$. The small deviations (up to a factor of $2$) from a constant $f_{\textrm{inst}}\eta_{e}$ line could result from cluster-to-cluster variations of $f_{\textrm{inst}}$, or from variations of the accretion flow across individual clusters (note that different instruments have different FOV).

\section{Other models}\label{sec:models}

In the preceding section we presented a simple model for the HXR emission from galaxy clusters, and have shown that it is consistent with the available HXR cluster data. In this section we consider alternative models, in which the HXR emission is predicted to be correlated with the cluster thermal emission. In such models, the HXR emission should be strongly dominated by emission from the cluster's core, which dominates the thermal X-ray emission. This is in contrast with the predictions of the model described in \S~\ref{sec:simple}, in which the HXR emission is produced at the cluster accretion shock. In the latter model, the HXR surface brightness is expected to be nearly uniform across the cluster, and enhanced along the (apparent) accretion shock ring (see eq.~\ref{eq:g_acc}).

A widely discussed model for the HXR emission, which predicts the HXR emission to be proportional to the thermal emission, is a model in which the relativistic electrons are secondaries produced by inelastic p-p collisions between cluster CRs and thermal intra-cluster gas \citep[e.g.,][]{dennison1980frh}. We therefore consider a simple version of this model in some detail below. In this simple version, the ratio between the CR energy density and the thermal energy density of the gas is constant. Although this ratio is predicted to scale with the gas density as $\rho^{-1/3}$ \citep{Jubelgas2008crf,pfrommer2008scr,kushnir2009non}, neglecting the radial dependence of this ratio is justified since the emission is dominated by the core \citep[see][for a detailed  discussion]{kushnir2009non}. As explained at the end of this section, our conclusion, that the data indicate that the spatial distributions of the HXR and of the thermal X-ray emission are different, is general and applies not only to the simplified secondary electron model.

In the secondary electron model, the ICM is assumed to contain a population of proton CRs with a power law energy distribution $\varepsilon^{2}dn_{CR}/d\varepsilon=\beta_{\textrm{core}}3nT/2$, where $n$ is the ICM number density and $\beta_{\textrm{core}}$ is the the ratio between the CR energy (per logarithmic particle energy interval) and the thermal energy. The HXR luminosity per logarithmic frequency interval is proportional in this model to the X-ray luminosity \citep[see][for details]{kushnir2009non},
\begin{eqnarray}\label{eq:IC to X}
\frac{\nu L_{\nu}^{\textrm{sec}}}{L_{X}}&\simeq&
1.1\cdot10^{-5}\beta_{\textrm{core},-4}T_{1}^{1/2}\frac{B_{\textrm{CMB}}^{2}}{B_{\textrm{CMB}}^{2}+B^{2}},
\end{eqnarray}
where $\beta_{\textrm{core},-4}=\beta_{\textrm{core}}/10^{-4}$ and $B$ is the intra-cluster magnetic field.  $B_{\textrm{CMB}}\equiv(8\pi aT_{CMB}^4)^{1/2}\approx3.2\,\mu \textrm{G}$ is the magnetic field, for which the magnetic energy density equals the energy density of the CMB. Note, that in contrast with the magnetic field at the accretion shock, which is much smaller than $B_{\textrm{CMB}}$, the magnetic field at the cluster core is expected to be $\ge B_{\textrm{CMB}}$.

The flux within a disk of angular radius $\theta$ for a cluster at a distance $d\sim cz/H_{0}$ is given \citep[for $\beta>0.5$, see][]{Sarazin1977xle} by
\begin{eqnarray}\label{eq:S for observations secIC}
F_{[\varepsilon_{1},\varepsilon_{2}]} (\theta) &=&
5.4\cdot10^{-15}h_{70}^2 \nonumber \\
&\times& T_{1}^{1/2}L_{X,45.5} \Lambda g_{\rm therm.}(\theta) \nonumber\\
&\times& \left\{%
\begin{array}{ll}
    B_{-5}^{-2}, & \hbox{$B\gg B_{\textrm{CMB}}$} \\
    0.1, & \hbox{$B\ll B_{\textrm{CMB}}$} \\
\end{array}%
\right.    \beta_{\textrm{core},-4} \,\frac{\textrm{erg}}{\textrm{cm}^{2}\,\textrm{s}},
\end{eqnarray}
where
\begin{eqnarray}\label{eq:g_thermal}
g_{\rm therm.}(\theta)&=& \left(3\beta-\frac{3}{2}\right) \left(\frac{z}{z_{\textrm{Coma}}}\right)^{-2}\nonumber \\
&\times&\int_{0}^{\min(\theta d/r_{c},r_{200}/r_{c})} \frac{\bar{r}d\bar{r}}{\left(1+\bar{r}^{2}\right)^{3\beta-1/2}}.
\end{eqnarray}
Here $L_{X,45.5}=h_{70}^{2}L_{X}/3\cdot10^{45}\,\textrm{erg}\,\textrm{s}^{-1}$, $\beta_{\textrm{core},-4}=\beta_{\textrm{core}}/10^{-4}$, and $B_{-5}=B/10\,\mu\textrm{G}$. In order to obtain an analytic relation between the luminosity and the surface brightness, we have assumed in deriving equations~\eqref{eq:S for observations secIC} and \eqref{eq:g_thermal} that the cluster emission extends to infinite radius. For $\beta$ values close to $0.5$, the derived relation deviates significantly from the one that would be obtained assuming that the emission is strongly suppressed beyond $r_{200}$. For clusters with $\beta$ close to $0.5$ we do not use, therefore, eq.~\eqref{eq:g_thermal}, but rather the relation obtained assuming emission is truncated beyond $r_{200}$.

\begin{figure}
\epsscale{1.2} \plotone{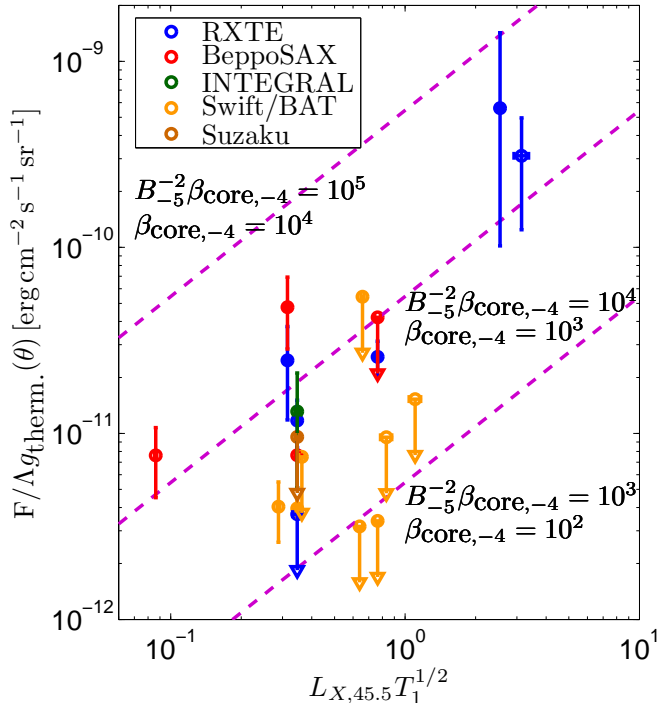} \caption{$F/\Lambda g_{\rm therm.}$ as function of $T_{1}^{1/2}L_{X,45.5}$. Dashed magenta lines show
constant values of $B^{-2}\beta_{\textrm{core}}$ (or $\beta_{\textrm{core}}$, see eq.~\ref{eq:S for observations secIC}).
\label{HXR_sec}}
\end{figure}

In the secondary model, $F/\Lambda g_{\rm therm.}$ scales as $T_{1}^{1/2}L_{X,45.5}$, with normalization that depends on $B^{-2}\beta_{\textrm{core}}$ (or only on $\beta_{\textrm{core}}$ if $B\ll B_{\textrm{CMB}}$). In figure~\ref{HXR_sec} we show $F/\Lambda g_{\rm therm.}$ as function of $T_{1}^{1/2}L_{X,45.5}$ (Vertical error bars reflect only the uncertainty in measured fluxes, $F$; we ignored the errors in the determination of $g_{\rm therm.}$, which are significantly smaller). Comparing the detected HXR emission with the constant $B^{-2}\beta_{\textrm{core}}$ (or $\beta_{\textrm{core}}$) lines of the figure, we find that $B_{-5}^{-2}\beta_{\textrm{core,-4}}=10^{4}$ (or $\beta_{\textrm{core,-4}}=10^{3}$ for $B\ll B_{\textrm{CMB}}$) is required in order to account for the detected fluxes. This implies that the total energy density of CR protons, $\log(\varepsilon_{\rm max}/\varepsilon_{\rm min})\times\beta_{\rm core}\approx20\beta_{\rm core}$, should exceed the thermal energy density of the gas in order for the emission from secondary electrons to account for the observed HXR emission. This strongly disfavors the secondary electron model. Moreover, the correlation between the radio and thermal X-ray emission of galaxy clusters suggests \citep{kushnir2009mfc} that the typical value of $\beta_{\textrm{core}}$ is $\simeq2\times10^{-4}$ and that $B\gg B_{\textrm{CMB}}$ for clusters which host radio halos.

Examining fig.~\ref{HXR_sec}, we find that the Swift/BAT detection and upper limits on $F/\Lambda g_{\rm therm.}$ are well below the values of $F/\Lambda g_{\rm therm.}$ inferred for clusters with similar $T$ from the detections of other instruments. The systematically lower HXR fluxes derived from the Swift measurements are naturally explained by the model described in \S~\ref{sec:simple}, in which the HXR emission is extended and not dominated by the cluster core. In this case, the lower Swift fluxes are due to its smaller FOV. Figure~\ref{HXR} demonstrates that this is a valid explanation (the upper limits on $F/\Lambda g_{\rm acc.}$ obtained by Swift are consistent with the values of $F/\Lambda g_{\rm acc.}$ inferred from the detection of other instruments). The lower HXR fluxes implied by Swift's observations may also be explained by assuming that the clusters observed by Swift are intrinsically different (e.g. having significantly lower values of $f_{\rm inst}\eta_e$) than those observed by other instruments. Since we have no reason to assume that the Swift clusters are systematically different than those observed by other instruments (all are prominently merging clusters), we conclude that the data disfavor all models, in which the spatial distribution of the HXR and of the thermal X-ray emission are strongly correlated.

\section{Discussion}\label{sec:conclusions}

We have presented in \S~\ref{sec:simple} a simple model, that explains the HXR emission from galaxy clusters as IC scattering of CMB photons by relativistic electrons accelerated in the accretion shock surrounding the cluster: The correlation predicted in this model between the HXR surface brightness and the cluster temperature is consistent with the observations, and the observed HXR luminosity is consistent with the fraction $\eta_{e}$ of shock thermal energy deposited in relativistic electrons being $\eta_{e}\sim0.1$ (see fig.~\ref{HXR_noup_cs}). The implied acceleration efficiency of electrons is similar to the acceleration efficiency of protons in the accretion shocks, which is inferred from the correlation between the radio flux and the thermal flux of galaxy clusters \citep{kushnir2009mfc}. The nonthermal luminosity and surface brightness produced by the accretion shock are determined in this model by the cluster thermal properties, and are given by eqs.~(\ref{eq:shock IC app}) and ~(\ref{eq:S for observations shockIC}).

Several comments are in place here regarding the estimated value of $\eta_{e}$. HXR observations do not allow one to determine $\eta_{e}$ directly. Rather, such observations constrain directly only the value of the product $\eta_{e}f_{\textrm{inst}}$, where $f_{\textrm{inst}}$ is the mass accretion rate measured in units of $M_{200}/t_H$ (see eq.~(\ref{eq:shock IC app})). We have found that the observed HXR fluxes are consistent with $\eta_{e}f_{\textrm{inst}}\sim0.1$. However, since the sample of clusters for which HXR observations are available is not complete, the inferred value of $\eta_{e}f_{\textrm{inst}}\sim0.1$ may be biased, i.e. may differ from its average value (over all clusters). In particular, since most of the clusters chosen for HXR observations are merging systems, in which enhancement of the accretion rate is expected \citep[see e.g.,][]{pfrommer2008scr}, the inferred value of $\eta_{e}f_{\textrm{inst}}$ is probably biased in the current sample towards values higher than average. Determination of the average value of $f_{\textrm{inst}}$ from numerical simulations and using a complete cluster sample, that may be produced by future HXR missions (e.g. NuStar, Simbol-X), would allow one to estimate the value $\eta_{e}$ more accurately.

Another limitation of the estimate of $\eta_e$ should be mentioned. We have assumed in our analysis that the ICM plasma is isothermal and in hydrostatic equilibrium. Deviations from this simple model near the virial radius may change our estimates for $\eta_{e}$. Although such deviations are only weakly constrained by observations, both observational \citep[e.g.][]{vikhlinin2005ctp} and theoretical \citep[e.g.][]{roncarelli2006sxr} analyses indicated that they are not large (for example, the accretion shock temperature is lower than the virial temperature by no more than a factor $\sim2$). Modifications of the ICM properties near the virial radius may be easily incorporated into our model. Improved (observational) determination of the ICM profile near the virial radius will therefore allow one to improve the accuracy of the determination of $\eta_{e}$.

Our model predicts the HXR surface brightness to be nearly uniform across the cluster, and enhanced along the (apparent) accretion shock ring (see eq.~\ref{eq:g_acc}). This is in contrast with models, in which the HXR emission is strongly correlated with the thermal X-ray emission, that is dominated by the cluster's core. We have shown in \S~\ref{sec:models} that the low values of HXR flux inferred from Swift's observations disfavor models, in which the HXR emission is dominated by the cores of the clusters: For an extended HXR emission, the low Swift fluxes are naturally explained as due to the smaller FOV of Swift, while for emission dominated by the cores of clusters the flux should not depend strongly on the FOV (see last paragraph of \S~\ref{sec:models} and compare figures~\ref{HXR} and~\ref{HXR_sec}). Moreover, it was shown in \S~\ref{sec:models} that a widely discussed model for HXR emission, in which the relativistic electrons producing the radiation are secondaries produced by inelastic p-p collisions between cluster CRs and thermal ICM \citep[e.g.,][]{dennison1980frh}, requires the total energy density of CRs to exceed the thermal energy density of the ICM in order to account for detected HXR fluxes.

Our model prediction, that the HXR emission is extended, may be tested by future HXR missions capable of producing high resolution HXR maps of clusters (NuStar and Simbol-X, e.g., may reach a resolution of tens of arcsec). This prediction could also be tested by future $\gamma$-ray observations. Our model predicts that cluster accretion shocks produce a $\gamma$-ray flux of (see eq.~\eqref{eq:S for observations shockIC})
\begin{eqnarray}\label{eq:S for observations shockIC2}
F_{\nu>\nu_{\min}}^{\textrm{IC,shock}}(\theta) &=&
4.7\cdot10^{-6}\left(\langle f_{\textrm{inst}}\rangle_{\theta}\eta_{e}\right)_{-1}\beta^{1/2} \nonumber \\ &\times& \left(\frac{f_{b}}{0.17}\right) T_{1}^{3/2} \left(\frac{\varepsilon_{\nu,\min}}{10\,\textrm{GeV}}\right)^{-1} \nonumber\\
&\times&  g_{\rm acc.}(\theta) h_{70}^2\,\textrm{ph}\,\textrm{cm}^{-2}\textrm{s}^{-1}.
\end{eqnarray}
This flux, which would have been marginally detectable by EGRET \citep[see detailed discussion for the Coma cluster in][]{kushnir2009non}, should be easily detectable by Fermi, which has a $\sim50$ times higher sensitivity and which may resolve the cluster (the angular resolution of Fermi reaches $0.1^{\circ}$ above $10\,\textrm{GeV}$, see http://www-glast.stanford.edu/).

Imaging Cerenkov telescopes may also detect the predicted nonthermal flux. However, it should be noted that flux predictions for energies $>1$~TeV are uncertain. $\sim1$~TeV photons are expected to be produced by the highest energy electrons accelerated in the accretion shocks \citep{loeb2000cgr}. Since, however, our estimate of the cutoff energy of the electrons is not robust \citep[see][for details]{keshet2003gri}, the $>1$~TeV flux may fall well below the prediction of eq.~(\ref{eq:S for observations shockIC2}). Lowering the energy threshold of the imaging Cerenkov telescopes to $\sim0.1\TeV$ would be very helpful in this context, since the prediction of eq.~(\ref{eq:S for observations shockIC2}) is more reliable at photon energies $\ll1$~TeV. It is important to mention here that measurements of the nonthermal emission in different energy bands (e.g. HXR and $\gamma$-rays) would allow one to constrain the energy distribution of the accelerated electrons.

A comment is appropriate regarding the synchrotron emission from the accretion shocks. Since the magnetic field at the accretion shock is expected to be weak, $\sim0.1\,\mu {\rm G}\ll B_{\rm CMB}$ \citep{waxman2000frb}, the synchrotron surface brightness produced by the accretion shock is negligible compared to that produced by secondaries \citep[e.g.,][]{kushnir2009non}. Detection of the accretion shock synchrotron emission is unlikely with present-day radio telescopes, but should be possible with next-generation telescopes such as the LOFAR and the SKA \citep[for detailed discussion, see][]{keshet2004iis}.

We finally note that our estimates for the IC flux cannot be used directly for the Soft X-ray band ($<1\,\textrm{keV}$), since the cooling time of the emitting electrons may exceed the dynamical time of the cluster (see eq.~\eqref{eq:t_cool}).

\acknowledgments This research was partially supported by AEC, Minerva and ISF grants. We thank the anonymous referee for comments that lead to significant improvements of the manuscript.



\bibliography{HXR}
\bibliographystyle{hapj}


\end{document}